\documentclass{article}

\newenvironment{anecdote}{\begin{list}{}{} \item}{\end{list}}

\newlength{\boxindent}
\newsavebox{\boxtext}
\newenvironment{sidebox}{\begin{lrbox}{\boxtext}
\begin{minipage}{\textwidth}\setlength{\parindent}{\boxindent}
\noindent}%
{\end{minipage}\end{lrbox}%
\begin{figure}\fbox{\usebox{\boxtext}}\end{figure}}
\newcounter{boxnum}
\newcommand{\boxcaption}[1]{\refstepcounter{boxnum}%
Box~\theboxnum: #1 \smallskip}

\begin{document}

\title{From Quantum Cheating to Quantum Security%
\footnote{This article originally appeared in Physics Today, vol.  53,
No. 11, p. 22, Nov. 2000. Also available online:
http://www.physicstoday.org/pt/vol-53/iss-11/p22.html
\copyright 2000, American Institute of Physics.  This article may be
downloaded for personal use only.  Any other use requires prior
permission of both the author and the American Institute of Physics.}
}

\author{Daniel Gottesman and Hoi-Kwong Lo}

\date{November 2000}

\maketitle

{\bf For thousands of years,
code-makers and code-breakers have been competing for
supremacy. Their arsenals may soon include a powerful
new weapon: quantum mechanics.}

Cryptography --- the art of code-making --- has a long history of
military and diplomatic applications, dating back to the
Babylonians. In World World Two, the Allies' feat of breaking the
legendary German code, Enigma, contributed greatly to the final
Allied victory. Nowadays, cryptography is becoming increasingly
important in commercial applications for electronic business and
electronic commerce. Sensitive data such as credit card numbers and
personal identification numbers (PINs) are routinely transmitted in
encrypted form. Quantum mechanics is a new tool for both
code-breakers and code-makers in their eternal arms race. It has
the potential to revolutionize
cryptography both by creating perfectly secure codes
and by breaking standard encryption schemes.

The most well-known application of cryptography is secure
communication~\cite{applied}. Suppose Alice would like to send a
message to Bob, but there is an eavesdropper, Eve, who is
wiretapping the channel. To prevent Eve from knowing the message,
Alice and Bob may perform encryption, i.e., transform the message
to something that is not intelligible to Eve during the
communication. On receiving the message, Bob inverts the
transformation and recovers the message (see
figure~\ref{fig:privacy}).


Bob's advantage over Eve lies in his knowledge of a secret,
commonly called the key, that he shares with Alice. The
key tells him how to decode the message. For example:

\begin{anecdote}
The rumble of Soviet tanks shook the Prague hotel room (number
117) as secret agent John Blond finished decoding his orders from
his superior N. He tore the used page from the codebook and
immediately burned it with his lighter.
\end{anecdote}

Blond is using a perfectly unbreakable cipher, a ``one-time pad.''
The secret codebook allows N and Blond to share a long secret
binary string --- the key --- before Blond leaves on his mission.
Whenever N would like to send a message to Blond, she first
converts it to binary.  She then takes the exclusive-OR (XOR)
between each bit of the message and the corresponding key bit to
generate the encrypted message, which is transmitted in a public
channel. The KGB can intercept the encrypted message, but without
the key, it is incomprehensible gibberish, offering no clue to the
contents of the original message. On the other hand, Blond, by
looking up the key in the codebook, can recover the original
message by taking the XOR between the encrypted message and the
key. Blond immediately burns the used page of the codebook to
prevent it from falling into the hands of the KGB in the future.

\section{Key Distribution Problem}

\begin{anecdote}
John Blond finally snapped shut the codebook and sighed. He had
been on duty in Czechoslovakia for so long that his codebook was
getting thin. He knew his days in Prague would soon be over: N
would have to recall him before he used up his whole codebook.
Blond recalled master cryptographer R's remonstration: ``This is
no joking matter, double-one seven. {\em Never} re-use the
one-time pad.''
\end{anecdote}

R was serious for a good reason. The
reuse of keys by the Soviet Union (due to the manufacturer's
accidental duplication of one-time-pad pages) enabled US
cryptanalysts to
unmask the atomic spy Klaus Fuchs in 1949~\cite{venona}.
When the key for a one-time pad is used more than once,
enemy cryptanalysts have the opportunity to look for patterns in
the encrypted messages which might reveal the key. Nevertheless,
excellent cryptosystems (known as symmetric cryptosystems)
have been developed that reuse
the key. The longer the key, the more secure the system.
For instance, a widely-used system is the Data Encryption Standard
(DES), which has a key length of $56$ bits. No method
substantially more efficient than trying all $2^{56}$ values of
the key is known for breaking DES. It is still conceivable,
however, that some yet unknown clever algorithm could defeat DES and its
cousins.

For top secret applications, therefore, the one-time pad is
preferable. Blond's predicament illustrates the drawback of the
one-time pad: when the secret key is used up, the code cannot be
used until the sender and receiver get together to share a new
secret key. Sending a courier with a new codebook into the Prague
Spring is a dangerous and unreliable business. Even if the courier
arrives, Blond and N can never be sure that the codebook was not
copied by the KGB during its journey.

This issue is known as the ``key distribution problem.'' A
possible solution is public key cryptography. Instead of a single
long key shared between the sender and receiver, public key
cryptography uses two sorts of keys: one public key, which is
known to the world, and one private key, known only to the
receiver. Anyone with the public key can send secret messages, but
only someone who knows the private key can read them. The
important defining feature of public key cryptography is that,
even knowing the encryption key, there is no known {\it
computationally efficient} way of working out what the decryption
key really is. As an example, the security of the most well-known
public key crypto-system, RSA, relies on the difficulty of
factoring large integers (see figure~\ref{fig:RSA}).


Public key cryptography can be used for another important task:
digital signatures. A digital signature exchanges the role of the
keys used in public key cryptography: the private key is used to
generate a signature, and the public key is used to verify it.
Only someone with the private key could have created the
signature.

\section{Quantum Code-breaking}

Both DES and RSA rely on an unproven assumption: there is no fast
algorithm to determine the secret key.  For instance, RSA is
believed to be secure because mathematicians throughout the world
have worked very hard to break it, steadily producing modest
improvements in factoring algorithms, but without groundbreaking
success.  By only modest increases in key size, users of RSA can
easily keep ahead even of the exponential growth in computing
power over the years.

Quantum mechanics changed this. In 1994, Peter Shor of AT\&T
invented a {\it quantum} algorithm for efficient factoring of
large numbers~\cite{shor}. The state of a quantum computer is a
superposition of exponentially many basis states, which correspond
to the states of a classical computer of the same size. By taking
advantage of interference and entanglement in this system, a
quantum computer can perform in a reasonable time some tasks that
would take ridiculously long on a classical computer. Shor's
discovery propelled the then obscure subject of quantum computing
into a dynamic and rapidly developing field, and stimulated scores
of experiments and proposals aimed towards building quantum
computers.

Another remarkable discovery was made by Lov Grover who in 1996
invented a quantum searching algorithm~\cite{grover}. To find one
particular item among $N$ objects requires checking $O(N)$ items
classically. With Grover's algorithm, a quantum computer need only
look up items $O \left(\sqrt N \right)$ times. It can be used to
radically speed up the exhaustive key search of DES (i.e., trying
all $2^{56}$ possibilities).

If a quantum computer is ever constructed in the future, much of
conventional cryptography will fall apart! The key lengths of
symmetric schemes like DES would have to be doubled due to
Grover's algorithm. The most commonly-used public key schemes are
RSA and classes based on discrete logarithms or elliptic curves;
Shor's algorithm breaks all of them. Even if it is decades until a
sufficiently large quantum computer can be built, this is a matter
of current concern: some data, such as nuclear weapons designs,
must still remain secret then, and it is important that today's
secret messages cannot be decoded tomorrow.

\section{Quantum Code-making}

Even if DES and RSA do fall apart, the one-time pad remains a
perfectly unbreakable cipher even against a quantum computer.
However, as discussed above, it has a serious catch --- the key
distribution problem: It presupposes that Alice and Bob share a
key that is secret and as long as the message. There is no way to
guarantee that in practice. Trusted couriers can be bribed or even
intercepted without their knowledge. More generally, classical
signals are distinguishable --- an eavesdropper can reliably
read the signals without changing them. Therefore, in classical
physics there is nothing, in principle, to prevent an eavesdropper
from wiretapping the key distribution channel passively.

Fortunately, quantum mechanics helps to make codes as well as
break them. (See \cite{book} and C. H. Bennett, ``Quantum
Information and Computation,'' Physics Today, Oct. 1995, p. 24.)
The Heisenberg
uncertainty principle dictates that it is fundamentally impossible
to know the exact values of conjugate variables such as the
momentum of a particle and its position. This apparent limitation
imposed by quantum mechanics can be used as a powerful tool in
catching eavesdroppers. The key idea is to use non-orthogonal
quantum states to encode information. More concretely, the essence
of quantum cryptography can be understood in a single question:
given a single photon in one of four possible polarizations
(vertical, horizontal, 45-degrees and 135-degrees), can one
determine its polarization with certainty? Surprisingly, the
answer is no. The rectilinear basis (the vertical and horizontal
directions) is conjugate to the diagonal basis (at 45 and 135
degrees), so the Heisenberg uncertainty principle forbids us from
simultaneously measuring both. More generally, experiments
distinguishing non-orthogonal states, even if only partially
reliable, will disturb the states.

The key distribution problem can be partially solved by quantum
mechanics using the idea of quantum key distribution (QKD).
The first and best-known protocol is usually called ``BB84''
because it was published in 1984 by Charles Bennett and
Gilles Brassard~\cite{bb84}. (See box~\ref{box:BB84} for details.)
In a prototypical quantum key distribution protocol, Alice sends
some nonorthogonal quantum states to Bob, who makes some
measurements.  Then, by talking on the phone (which need not
be secure), they decide if Eve has tampered with the quantum
states. If not, they have a shared key, which is guaranteed to
be secret. Note that Alice and Bob must share some authentication
information to begin with: otherwise, Bob has no way to know that
he is really talking to Alice on the phone, and not a clever
mimic. The key generated by QKD can subsequently be used for both
encryption and authentication, thus
achieving two major goals in cryptography.


\begin{sidebox}
\boxcaption{BB84} \label{box:BB84}

In the most well-known quantum key distribution
scheme, BB84, Alice sends Bob a sequence of photons, each independently
prepared in one of four polarizations (horizontal, vertical, $45$
degrees, or $135$ degrees). For each photon, Bob randomly picks
one of the two (rectilinear and diagonal) bases to perform a
measurement. He keeps the measurement outcome secret. Now Alice
and Bob publicly compare their bases. They keep only the
polarization data for which they measured in the same basis. In
the absence of errors and Eve, these data should agree. (They
throw away the data for which their bases disagree.)

To test for tampering, they now choose a random subset of the
remaining polarization data and publicly announce them. From there
they can compute the error rate (i.e., the fraction of points
where their values disagree).
If the error rate is unreasonably high, above around say 10\%,
they throw away all the data (and perhaps try again later).
If the error rate is acceptably small, they perform error
correction and also ``privacy amplification'' to distill out a
shorter string which will act as the secret key. These steps
essentially ensure that their keys agree, are random, and are
unknown to Eve.

Other quantum key distribution schemes have also been proposed.
For example, Artur Ekert (of Oxford) proposed one based on quantum
mechanically correlated (i.e., entangled) photons and using Bell
inequalities as a check of security. In 1992,
Charles Bennett of IBM proposed a simple quantum key distribution
scheme (called B92) which uses only two non-orthogonal states.

\end{sidebox}

\section{Experimental quantum key distribution}

Quantum key distribution (QKD) is an active experimental subject.
%
%
The first working experimental prototype was constructed in 1989
at IBM, Yorktown Heights \cite{32cm}. It transmitted quantum signals over
32~cm of free air. Since then, various groups including those led by
Paul Townsend of BT, Jim Franson of John Hopkins University,
Nicolas Gisin and Hugo Zbinden of the University of Geneva, and
Richard Hughes of Los Alamos National Labs have made important
contributions.  A primary focus has been a series of impressive
experiments over commercial optical fibers.  The world record
distance for QKD, at the time of writing, is about 50km~\cite{50km}.
One of the long distance experiment was performed by Los Alamos National
Laboratories and is depicted in figure~\ref{fig:QKD}.

Most experiments to date have used variants of either the BB84 or B92
schemes (see box~\ref{box:BB84}), although recently three groups ---
one led by Paul Kwiat of Los Alamos, Gisin and Zbinden's group at
Geneva, and a collaboration led by Anton Zeilinger of the University
of Vienna and Harald Weinfurtur of the University of Munich --- have
independently implemented EPR-based protocols.  In the BB84 and B92
schemes, typically a single-photon source is simulated using
attenuated coherent states --- on average, only a fraction of a photon
is actually sent.  When taken in conjunction with absorption in the
fiber, only a small fraction of arriving laser pulses actually contain
a photon.  This does not interfere much with key distribution, since
only the photons that reach Bob are used in the protocol.  The key is
generally encoded in either the polarization or the phase of the
photon.  Error rates in the photons actually received are usually a
few percent.

For commercial applications in say a local area network environment,
it is useful for a quantum cryptographic system to be integrated
into a passive multi-user optical fiber network and its
equipment to be miniaturized.
Paul Townsend's group at BT (now of Corning) has done much work in
this area~\cite{townsend}. As for point-to-point applications,
the Geneva group
has devised a so-called ``plug and play'' system which automatically
compensates for polarization fluctuations~\cite{plug}.
Such systems might
someday convey secret information between government agencies
around Washington D.C., or connect bank branches within a city.

Quantum key distribution has
also been performed in open air, during daylight, with a current
range of about 1.6~km~\cite{air}. An even more ambitious proposal
to perform a ground to satellite quantum key distribution
experiment has been made by John Hopkins University and, more
recently and in more detail, by Los Alamos National Laboratories.
If successful, quantum cryptography may be used to ensure the
security of command control of satellites from control centers on
the ground.

Future experiments will aim to make quantum key distribution
more reliable, to integrate it with today's communications
infrastructure,
and to increase the distance and rate of key
generation.  Another ambitious goal would be to produce a
quantum repeater using techniques of quantum error correction.
Such an accomplishment will require substantial technical
breakthroughs, but would allow key distribution over
arbitrarily long distances.


\section{Is QKD Secure?}

While experiments in quantum key distribution forged ahead, the
theory developed more slowly. A clever Eve can adopt many possible
strategies to fool Alice and Bob, including subtle quantum attacks
entangling all of the particles sent by Alice. Taking all
possibilities into account, as well as the effects of realistic
imperfections in Alice and Bob's apparatus and channel, has been
very difficult. A long series of partial results by many people
has appeared over the years, addressing restricted sets of
strategies by Eve~\cite{earlier}, but only in the past few years
have complete proofs appeared.

One class of proofs, by Dominic Mayers~\cite{mayersqkd} and
subsequently others~\cite{others} attack the problem directly, and
prove the security of the standard BB84 protocol. Another
approach, by one of us (HKL) and H.~F.~Chau~\cite{qkd}, instead
proves the security of a new protocol using quantum
error-correcting codes \cite{book}. (For information on
quantum error correction, see
J. Preskill, ``Battling Decoherence: The Fault-Tolerant
Quantum Computer'', Physics Today, June 1999, p. 24.)
This approach allows one to
apply {\it classical} probability theory to tackle a {\it quantum}
problem directly. It works because the relevant observables under
consideration all commute with each other.
While conceptually
simpler, this protocol
requires a quantum computer to implement. The two classes of
approaches have been unified by Peter Shor and John
Preskill~\cite{shorpre}, who showed that a quantum code protocol
could be modified to become BB84 without compromising its
security.

The proof of the security of QKD is a fine theoretical result, but
it does not mean that a real QKD system would be
secure \cite{norbert}. Some
known and unknown security loopholes might prove to be fatal.
Apparently minor quirks of a system can sometimes provide a lever
for an eavesdropper to break it. For instance, instead of
producing a single photon, the laser may produce two; Eve can keep
one and give the other to Bob. She can then learn what
polarization Alice sent without revealing her presence. There are
various possible solutions to this particular problem; it is the
unanticipated flaws that present the greatest security hazard.
Ultimately, we cannot have confidence that a real-life quantum
cryptographic system is secure until it has stood up to attacks
from determined real-life adversaries. Traditionally, breaking
cryptographic protocols has been regarded as important as making
them --- the protocols that survive are more likely to be truly
secure. The same standard will have to be applied to quantum key
distribution.

\section{Post-Cold-War Applications}

There are many problems beyond secure communication that can be
addressed by cryptography.

\begin{anecdote}
Alice and Bob are considering going on a date, but neither is
willing to admit their interest unless the other is also
interested. How can they decide whether or not to date without
letting slip any unnecessary information?
\end{anecdote}

This dating problem can be phrased as the problem of computing a
function $f(a,b) = ab$, where $a$ and $b$ are single bits held by
Alice and Bob ($0$ = not interested, $1$ = interested). Problems
like this can be solved classically using variants of public key
cryptography, which we know might be rendered insecure by quantum
computers. By exchanging quantum states, can Alice and Bob solve
the above dating problem with absolute security?

There are many possible functions $f$ that two people might wish
to compute together, too many to address all of them individually.
Instead, cryptographers rely on a suite of primitive operations
that can be composed to build more complex functions. One
important protocol is called bit commitment, and it is the
electronic equivalent of a locked box: Alice chooses a bit, $0$ or
$1$, and writes it on a piece of paper, which she deposits in the
box. She gives the box to Bob, but keeps the key. She cannot
change what she wrote, and without the key, Bob cannot open the
box, but at some later point, Alice can give Bob the key and
reveal her bit. By itself, bit commitment is useful mostly for
debunking professional psychics, but it serves as a useful
building block for more interesting functions.

Consider the following bit commitment scheme proposed by Bennett
and Brassard~\cite{bb84}: if Alice wishes to commit to a $0$, she
sends Bob a polarized photon in the rectilinear basis; if she
wishes to commit to a $1$, she sends Bob a polarized photon in the
diagonal basis. In either case, Alice flips a coin to decide which
of the two polarizations to send. Bob has no way to tell which
basis Alice used --- either way, his state is perfectly random.
When Alice unveils her bit, she tells Bob which of the four states
she used. Then Bob can measure in the appropriate basis to verify
that Alice is telling the truth. If she lies about which basis she
used, Bob has a $50\%$ chance of finding out. If the protocol is
repeated many times, Alice's chance of successfully cheating is
abysmally small.

This protocol is secure against a classical cheater, who does not
have much ability to store and manipulate quantum states. But as
Bennett and Brassard recognized, a {\em quantum} cheater can break
the protocol. Suppose, instead of picking a specific state and
sending it to Bob, Alice instead creates an entangled pair of
photons, $ ( | HV \rangle - | VH\rangle ) / \sqrt{2}$ (this is
also called an Einstein-Podolsky-Rosen, or EPR, pair), and sends
the second photon to Bob, keeping the first one. She stores the
quantum state of the first photon and delays measuring it. Suppose
that when the time comes for Alice to open the commitment, she
decides she would like the committed bit to read $0$, which
requires her to specify a state in the rectilinear basis. Alice
knows that if she and Bob measure in the same basis, they will get
opposite results. Therefore, she can measure her photon in the
rectilinear basis and tell Bob he has the opposite polarization,
and she will always be right.

If Alice instead wishes the committed bit to read $1$, she needs a
state in the diagonal basis.  But
\begin{equation}
( | HV \rangle - | VH\rangle ) / \sqrt{2} = ( | 45^\circ,
135^\circ \rangle - |135^\circ,  45^\circ \rangle / \sqrt{2}
\end{equation}
She measures her particle in the diagonal basis, and can again be
sure that Bob's measurement outcome will be opposite to hers.
Quantum cheating allows Alice to change her mind at the last
minute without being caught by Bob, thus totally defeating the
purpose of bit commitment.

Nonetheless, more sophisticated schemes for quantum bit commitment
were proposed, and for a long time were believed to be secure.
Eventually, the bubble burst and it was shown
that the quantum cheating strategy discussed
above --- which uses the EPR effect and delayed measurements --- can
be generalized to break all two-party quantum bit commitment
schemes~\cite{bit}: If Alice and Bob hold one of two pure quantum
states which are indistinguishable to Bob, then Alice, acting
unilaterally, can change one to the other. Therefore, the two
basic requirements of bit commitment --- that Bob does not know
the bit and that Alice cannot change it --- are fundamentally
incompatible with quantum mechanics.

The strength of the proof lies in its generality.
The idea is to treat the whole system as if it were quantum
mechanical, extending the part that was originally quantum to
include any dice, measuring devices, and classical computation that
appear in the protocol. For this point of view, the original
protocol is equivalent to a purely quantum one, with some of
the outputs being thrown in the trash
(see figure~\ref{fig:church}).  Note
that throwing something away can never help a cheater, so
we might as well assume that the state shared by Alice and Bob is
completely determined by the protocol --- it is a pure quantum
state, such as an EPR pair.
That substantially reduces the complexity of the problem; it is
not difficult to show that when Alice and Bob
hold a pure state, quantum bit commitment is impossible.


Following the fall of quantum bit commitment, other important
basic quantum cryptographic protocols have also been proven to
be insecure by one of us (HKL), thus
leaving the field in shambles. What is
left? Some problems are too similar to bit commitment, and cannot
be done at all.

Other problems have more modest goals, and {\em can} be solved by
quantum protocols.  For instance, Lior Goldenberg, Lev Vaidman,
and Stephen Wiesner of Tel Aviv University have proposed a method
of ``quantum gambling,'' where a cheater must pay a large fine
whenever he is caught. The majority of problems lie in a middle
ground --- we do not know whether or not they are possible. The
dating problem is an example. Many approaches to it tread too near
bit commitment, and are doomed to failure, but it's possible there
are others, as yet undiscovered, which do not.

\begin{sidebox}
\boxcaption{Glossary}
\label{box:glossary}

\begin{description}

\item[B92] a quantum key distribution scheme using two nonorthogonal states.

\item[BB84] a quantum key distribution scheme.  See box~\ref{box:BB84}.

\item[cryptanalysis] the art of code-breaking

\item[cryptography] the art of code-making

\item[Data Encryption Standard (DES)] a US federal standard symmetric
                             encryption scheme for unclassified but
                             sensitive data.

\item[key] a random string of numbers employed in encryption or decryption

\item[one-time pad] a perfectly secure encryption scheme in which the
              key is as long as the message and never re-used.

\item[private key] a decryption key that is kept secret.

\item[public key] an encryption key that is publicly announced.

\item[public key cryptography] see figure 2.

\item[Rivest-Shamir-Adleman (RSA) scheme] see figure 2

\item[quantum key distribution (QKD)] a scheme based on quantum mechanics
                         that allows two users to share a common
                         string of secret numbers.

\item[symmetric cryptosystem] a cryptographic system with
the same private key for both encryption and decryption.

\end{description}

\end{sidebox}

\section{Physics Today, Cryptology Tomorrow}

Quantum computers are still on the drawingboards, and quantum
cryptography systems are only prototypes. Still, there are a number of
reasons for thinking about quantum cryptology today.  Unlike other
cryptosystems, the security of quantum key distribution is based on
fundamental principles of quantum mechanics, rather than unproven
computational assumptions.  QKD eliminates the great threat of
unanticipated advances in algorithms and hardware breaking a
widely-used cryptosystem.  Small-scale QKD systems are well within the
capabilities of today's technology, and commercial systems could be
available within a few years (although whether such systems are widely
adopted depends on many non-academic factors such as cost. For further
discussion, see H.-K. Lo, quant-ph/9912011).

Furthermore, grappling with the problems posed by quantum protocols
can give us insight into more general questions about quantum
mechanical systems in many fields of physics.
For instance, one reason it is hard to analyze protocols and
attacks is that they frequently involve a combination of quantum
and classical behaviors. In considering bit commitment in
the last section, it was
possible to replace classical parts of the protocol with a quantum
description, an approach which is useful for many problems inside
and outside the field of quantum cryptography. It is sometimes
called the Church of the Larger Hilbert Space, following John
Smolin of IBM --- all quantum operations, including measurements,
are unitary when considered as acting on a larger Hilbert space
(figure~\ref{fig:church}).

Finally, quantum mechanics changes the world of cryptology, and it
is important to know what the new terrain will look like in order
to decide on cryptographic standards that may last for decades.
In a world where quantum computers and communication are commonplace,
today's most widespread public key cryptosystems would no longer
work; in the worst case, perhaps {\em no} public key cryptosystem
will work. If so, symmetric cryptosystems and quantum key
distribution would partially fill the gap, allowing secure
communication. Unfortunately, digital signatures would fail as
well, meaning important communications would need to be notarized
by a trusted third party.

Of course, quantum key distribution and symmetric cryptosystems
are not useful in situations when Alice and Bob have never met
before. Solving this problem would probably require a quantum
cryptographic center, which could verify the identity of both of
them. The center would have to be known and trusted by both Alice
and Bob.

Problems beyond secret communication and digital signatures are a
mixed bag. Many, such as bit commitment and perhaps the dating
problem, would be impossible, whereas others, such as quantum
gambling, could be carried out with complete security.

This is just one of a number of possible futures: perhaps some
new or existing
public key cryptosystems will survive quantum computation, or
perhaps there will be new public key systems that can only run on
a quantum computer. Perhaps quantum computers will always remain
difficult to build (we believe that this is unlikely),
and public key cryptography will remain
widespread, despite its potential flaws.  Only
time will tell who benefits more from quantum cryptology: the
code-makers or the code-breakers.

\begin{sidebox}
\boxcaption{Decoding the message in Figure~\ref{fig:privacy}}

\label{box:puzzle}
The code is a ``Caesar's cipher,'' where each
letter is shifted by a fixed number of places in the alphabet.
In this case, the shift is three places.
\end{sidebox}

\begin{sidebox}
\boxcaption{About the Authors}
\label{box:authors}

Daniel Gottesman is a postdoctoral fellow in the Theory Group of
Microsoft Research, Redmond, Washington. Hoi-Kwong Lo is Chief
Scientist and Senior Vice President, Research and Development of MagiQ
Technologies, Inc., New York, NY, http://www.magiqtech.com, a start-up
company that focuses on the commercial exploitation of quantum
information technology.
\end{sidebox}

\pagebreak

\begin{figure}
\caption{Secure communication: a) Alice sends a message to Bob
through a communication channel, but an eavesdropper, Eve, is
wiretapping. b) A message is encrypted by Alice using an
encryption key. The encrypted message, the {\em ciphertext}, is
now unintelligible to Eve. Bob, who has the same key as Alice, can
decrypt the ciphertext and recover the original message.
(The code used in this figure is not very secure. Try
breaking it yourself; the solution is at the end of the article.)}

\label{fig:privacy}
\end{figure}

\begin{figure}
\caption{The RSA public key crypto-system: The most well-known
public-key system is called RSA, after its inventors Ronald
Rivest, Adi Shamir and Leonard Adleman. It is based on modular
arithmetic over a large base $N$, which is the product of two
large primes $p$ and $q$. If $x$ is relatively prime to $N$, the
Euler-Fermat theorem tells us that $x^r \equiv 1 \bmod N$, where
$r=(p-1)(q-1)$. The public key is a pair of numbers $(N, e)$, and
the private key is $d$, with $ed \equiv 1 \bmod r$ (i.e., $ed = kr
+1$ for some integer $k$). To encrypt a message $m$, a sender
Alice computes $y \equiv m^e \bmod N$. To decrypt the message $y$,
the receiver Bob computes $y^d \bmod N \equiv m^{ed} \equiv m$.
Note that, to do so, Bob has to know the private key $d$. Anyone
can send Bob an encrypted message, but only Bob can decrypt it.}
\label{fig:RSA}
\end{figure}

\begin{figure} 
\caption{Experimental quantum key distribution: 
a) Schematic of the 48~km optical fiber quantum 
key distribution experiment at Los Alamos~\cite{50km}, 
which implements the B92 two-state protocol.  The laser 
with wavelength 1.3 $\mu$m is first attenuated to approximate 
a single-photon source, which will ultimately produce the
key bits.  The signal generated 
is then passed through Alice's interferometer. 
The two nonorthogonal states of the B92 protocol are 
realized as two possible settings for the phase delay, $\phi_A$, 
in one branch of the interferometer.  To measure the 
state, Bob passes the photon through his own interferometer, 
adding one of two possible phase shifts, $\phi_B$, and
detects the photon in one of the two cooled detectors,
depending on the choice of both $\phi_A$ and $\phi_B$.
The room temperature detector looks for a bright pulse 
from a second laser with wavelength 1.55 $\mu$m,
which tells Bob when to 
expect a photon from Alice.  Air gaps in both 
interferometers are used for adjustments to keep 
Alice and Bob properly synchronized. 
b) Photo of the Los Alamos experiment. 
Each signal originates with Alice's computer, passes 
through the 48~km of optical fiber, and returns to 
Bob's computer, just next to Alice's. 
(Figure courtesy of Los Alamos National Labs.)} 
\label{fig:QKD} 
\end{figure}

\begin{figure}
\caption{The Church of the Larger Hilbert Space.
The most general operation compatible with quantum mechanics,
including measurement and generating random numbers,
can be described using three steps.  First, augment the system
with an additional system.  Second, perform a unitary
transformation involving the original system and the new system.
Third, throw away some of the outputs.  This picture applies
equally well to a long series of operations, which can all be
pushed into the second step of unitary transformations.  This
step is purely quantum, allowing us to delay the complications
of measurement until the end of the evolution.  In the figure,
the horizontal lines represent the paths of elementary
quantum subsystems, such as two-level quantum bits (``qubits''),
which enter on the left and interact via the unitary
transformation $U$.}
\label{fig:church}

\end{figure}

\end{document}